\documentclass[
  aip,
  jcp,          
  reprint,
  superscriptaddress,
  longbibliography
]{revtex4-2}

\usepackage{amsfonts,amscd,mathrsfs,amsmath,amsthm,amssymb}
\usepackage{mathtools}
\usepackage{xparse}
\usepackage{graphicx}
\usepackage{float}
\usepackage{cancel}

\usepackage{youngtab}
\usepackage{mhchem}

\PassOptionsToPackage{hyphens}{url}\usepackage{hyperref}

\usepackage[capitalise]{cleveref}
\usepackage{physics}
\usepackage{bm}
\usepackage{bbm}
\usepackage{caption}

    \allowdisplaybreaks

    \newcommand{\CC}{\mathbb{C}} 

    \DeclareMathAlphabet{\mathsfit}{T1}{\sfdefault}{\mddefault}{\sldefault}
    \SetMathAlphabet{\mathsfit}{bold}{T1}{\sfdefault}{\bfdefault}{\sldefault}

    \newcommand{\SU}{\mathrm{SU}}

    \newcommand{\U}{\mathrm{U}}

\newcommand{\HH}{\mathcal{H}}

\usepackage{xcolor}

\begin{document}

\title{Nuclear–Spin Statistical Weights from Young Diagrams}
\thanks{These authors contributed equally to this work.\\ekubischta@fsu.edu, iteixeira@ucsd.edu}

\author{Eric Kubischta}
\affiliation{Quantum Initiative, Florida State University, Tallahassee, FL 32306}
\affiliation{Department of Mathematics, Florida State University, Tallahassee, FL 32306}
\author{Ian Teixeira}
\affiliation{Department of Mathematics, University of California, San Diego, CA 92093}

\begin{abstract}
Nuclear–spin statistical weights and selection rules in molecular spectroscopy are governed by the
permutation symmetry of identical nuclei, although rigid and semi-rigid molecules typically realize
only a proper subgroup of the full symmetric group.
Building on the Schur–Weyl framework of Schmiedt, Jensen, and Schlemmer, we extend this approach to
molecular geometries whose rovibrational motion realizes cyclic and dihedral permutation subgroups. By combining the Schur–Weyl decomposition of the $n$-spin Hilbert space with the
Kraśkiewicz–Weyman–Adin–Roichman major-index branching rule for the restriction
$S_N \downarrow C_m$, we obtain a fully combinatorial method for determining nuclear–spin symmetry
species and statistical weights.
Each standard Young tableau corresponds to a definite cyclic character determined by its major
index, with reflections yielding the associated dihedral symmetry, so that nuclear–spin species
follow directly from tableau data without projection operators or case-specific constructions. Applications to \ce{XeOF4}, \ce{SF4}, benzene, deuterated benzene, and the tropylium cation demonstrate that
the method applies uniformly to realistic molecular point groups and reproduces established
statistical weights while providing a transparent spin-resolved structure.
The approach supplies the symmetry information required for rovibrational line-intensity modeling
and provides a practical, automatable framework for nuclear–spin symmetry analysis in polyatomic
molecules.
\end{abstract}

\maketitle 

\section{Introduction}

Permutation symmetry of identical nuclei is essential for the interpretation of molecular spectra.  In standard Born--Oppenheimer theory the electronic structure
determines an effective potential energy surface, and the subsequent nuclear-motion
problem yields rovibrational wavefunctions.  Neglecting the small hyperfine
contribution to the energy, the nuclear-spin degrees of freedom factor from the
spatial motion, but they do not decouple from symmetry: Fermi--Dirac or
Bose--Einstein statistics restrict the rovibrational states that can be populated and
fix their relative nuclear-spin statistical weights. These weights are essential for interpreting intensity patterns,
establishing assignments in high-resolution spectra, and describing
low-temperature kinetics in systems where spin conversion is slow.
More generally, they determine the statistical factors attached to
rovibrational symmetry species and therefore enter the analysis of
spectral line intensities, nuclear-spin isomer populations, and
symmetry selection rules in reactions involving identical nuclei.

For a small number of identical nuclei one can determine the nuclear--spin symmetry
species by explicit spin coupling or by direct symmetry projection.  As $N$ grows,
however, the bookkeeping quickly becomes cumbersome, and the standard tools of the
field---character tables, projection operators, and cycle-index formulas---tend to
obscure rather than reveal the organizing structure.  Moreover, for rigid and
semi-rigid molecules the \emph{full} symmetric group $S_N$ of all permutations of the
identical nuclei is rarely realized physically: only those permutations induced by
feasible molecular motions contribute.  The symmetry relevant for spectroscopy is
therefore a proper subgroup of $S_N$ determined by the molecular geometry. Determining nuclear--spin statistical weights therefore reduces to
restricting the $S_N$ representations arising in the Schur--Weyl
decomposition of the nuclear--spin space to the permutation subgroup
realized by the molecular motions. 

Even when hyperfine effects are neglected and the nuclear--spin and rovibrational
energy contributions decouple to excellent approximation, the symmetry constraint
does not: the total molecular wavefunction must transform with the required
permutation symmetry of the identical nuclei.  Consequently, a rovibrational
symmetry species $\Gamma_{\mathrm{rovib}}$ admits only certain nuclear--spin species
$\Gamma_{\mathrm{nspin}}$, and the number of compatible spin functions is the
corresponding nuclear--spin statistical weight.  These weights determine which
rovibrational symmetry species occur with nonzero statistical weight and underlie
the familiar intensity ratios between symmetry-distinct rotational and rovibrational
lines.

A major conceptual advance was provided by Schmiedt, Jensen, and Schlemmer
\cite{SchmiedtJensenSchlemmer2016}, who reformulated nuclear--spin statistics using
Schur--Weyl duality.  In their framework the nuclear--spin Hilbert space of $N$
identical spin-$I$ nuclei,
\[
\HH_d^N \;=\; (\CC^{2I+1})^{\otimes N}, \qquad d=2I+1,
\]
is treated as a representation of $\U(d)\times S_N$.  Schur--Weyl duality \cite{fultonharris} yields a
joint decomposition into irreducible representations of the unitary (and hence
rotational) symmetry acting on each single-spin space and of the permutation symmetry
of the nuclei, making precise the sense in which angular-momentum and permutation
classifications are coupled rather than independent.  This viewpoint resolves the difficulties that arise when attempting to reconcile the $ S_N $ based methods of \cite{LonguetHiggins1963NonRigid,Quack1977SymmetrySelection}  with the $ \SU{2} $ based methods of \cite{ParkLight2007SpinSelection,Oka2004NuclearSpinSelection} as applied to the problem of analyzing the permutation symmetry
and total-spin structure for multiple nuclei with $I>\tfrac12$. Subsequent work has reinforced the usefulness of this perspective, for example Bengs
emphasized the role of combined rotational--permutational structure in stabilizing
long-lived nuclear-spin order \cite{Bengs2020}. 

Despite this progress, a practical gap remains between the Schur--Weyl decomposition
under $S_N$ and the symmetry reductions required in high-resolution spectroscopy of
rigid molecules.  In such applications the physically realized permutation symmetry is
the \emph{permutation projection} of the molecular point group, i.e.\ the subgroup of
$S_N$ generated by symmetry-allowed nuclear motions.  For many common geometries this
subgroup contains an $m$-fold rotation about a principal axis and is therefore cyclic
(or, after adjoining a reflection, dihedral).  While Schur--Weyl duality provides a
powerful organization of $\HH_d^N$ as an $S_N$-module, it does not by itself give a
general, systematic rule for restricting these $S_N$ representations to the cyclic
subgroups that arise directly from molecular geometry.

The purpose of the present work is to supply this missing step in a form that is both
transparent and computationally convenient.  Starting from the Schur--Weyl
decomposition of $\HH_d^N$ into $\U(d)\times S_N$ irreducibles, we exploit a purely
combinatorial branching rule for the restriction
\[
S_N \downarrow C_m,
\]
due to Kra\'skiewicz--Weyman and Adin--Roichman
\cite{KraskiewiczWeyman1989,AdinRoichman2001}.  This theorem states that, upon
restricting an $S_N$ irrep to a cyclic subgroup generated by an element of order $m$,
the multiplicity of the one-dimensional $C_m$ character $\chi_j$ is determined by the
number of standard Young tableaux whose major index is congruent to $j \pmod m$.
In this way, the abstract $S_N$ classification furnished by Schur--Weyl duality is
converted directly into the cyclic species relevant for rigid molecular geometries.

The resulting procedure avoids explicit character-theoretic calculations.  Once the
standard Young tableaux of the relevant shapes are enumerated, the cyclic species and
their multiplicities follow immediately from the major index, without recourse to
projection operators or cycle-index polynomials.  When the physical point group is
dihedral (or contains reflections), the cyclic information recombines into point-group
irreducible representations after adjoining a single reflection generator, so that the
same tableau data yield both nuclear--spin statistical weights and a refined,
spin-resolved symmetry classification suitable for rovibrational applications.

\section{Preliminaries}

The framework developed in this work rests on two standard pieces of
representation theory:
(i) the Schur--Weyl decomposition of the nuclear--spin Hilbert space into
irreducible representations of $\U(d)\times S_N$, and
(ii) the parametrization of these irreps by Young diagrams and tableaux.
We summarize the essential ingredients here, emphasizing cases of direct
spectroscopic relevance.
These decompositions provide the input data for the cyclic-subgroup branching
rules where each standard Young tableau contributes a definite
$C_m$ species via its major index.

\subsection{Young diagrams}

For a system of $N$ identical nuclei, the full permutation symmetry is described
by the symmetric group $S_N$.
Its irreducible representations are indexed by Young diagrams with $N$ boxes.
For example, the five possible Young diagrams with $ N=4 $ boxes are
\[
{\tiny\Yvcentermath1\yng(4)},\quad
{\tiny\Yvcentermath1\yng(3,1)},\quad
{\tiny\Yvcentermath1\yng(2,2)},\quad
{\tiny\Yvcentermath1\yng(2,1,1)},\quad
{\tiny\Yvcentermath1\yng(1,1,1,1)},
\]
and these classify the possible permutation symmetries of four identical nuclei, such
as the fluorines in \ce{XeOF4} or \ce{SF4}.
Although the full $S_N$ symmetry is rarely realized physically in rigid
molecules, its decomposition provides the starting point for the cyclic
restriction $S_N \downarrow C_m$.

For $I=\tfrac12$ ($d=2$), only Young diagrams with at most two rows appear:
\[
{\tiny\Yvcentermath1\yng(4)},\quad
{\tiny\Yvcentermath1\yng(3,1)},\quad
{\tiny\Yvcentermath1\yng(2,2)}
\]
Counting semistandard tableaux of these shapes using the alphabet $\{1,2\}$
yields the classical decomposition
\begin{equation}
(\CC^{2})^{\otimes 4}
\;\cong\;
5\,{\tiny\Yvcentermath1\yng(4)}
\;\oplus\;
3\,{\tiny\Yvcentermath1\yng(3,1)}
\;\oplus\;
1\,{\tiny\Yvcentermath1\yng(2,2)}.
\label{eq:spinhalf}
\end{equation}
This case underlies many systems of interest, including the four fluorine spins
in \ce{XeOF4} and \ce{SF4}, which will be analyzed below.

\subsection{Standard Young tableaux }

To determine the \emph{structure} of each $S_N$ irrep, we use
\emph{standard Young tableaux} (SYTs): labelings of the Young diagram with
$\{1,\dots,N\}$ increasing strictly along rows and down columns.
The number of such tableaux equals the dimension of the $ S_N $ representation, but more importantly
for this work, each tableau contributes one basis vector whose major index
controls its transformation under a cyclic permutation.

For $N=4$ the standard tableaux of each partition are:
\begin{align*}
    & {\tiny\Yvcentermath1 \young(1234) } \\
    & {\tiny\Yvcentermath1 \young(134,2) } \quad
      {\tiny\Yvcentermath1 \young(124,3) } \quad
      {\tiny\Yvcentermath1 \young(123,4) } \\
    & {\tiny\Yvcentermath1 \young(12,34) } \quad
      {\tiny\Yvcentermath1 \young(13,24) } \\
    & {\tiny\Yvcentermath1 \young(12,3,4) } \quad
      {\tiny\Yvcentermath1 \young(13,2,4) } \quad
      {\tiny\Yvcentermath1 \young(14,2,3) } \\
    & {\tiny\Yvcentermath1 \young(1,2,3,4) } 
\end{align*}

The number of tableaux of each shape is summarized in
Table~\ref{tab:irrepdim}.
These counts reappear throughout both the Schur--Weyl decomposition and the cyclic-branching analysis developed below.

\begin{table}[htp]
    \centering 
    \begin{tabular}{|c|c|c|c|c|c|} \hline 
        Irrep &  $\scalebox{0.5}{\tiny\Yvcentermath1\yng(4)}$ &
        $\scalebox{0.5}{\tiny\Yvcentermath1\yng(3,1)}$ &
        $\scalebox{0.5}{\tiny\Yvcentermath1\yng(2,2)}$ &
        $\scalebox{0.5}{\tiny\Yvcentermath1\yng(2,1,1)}$ &
        $\scalebox{0.5}{\tiny \Yvcentermath1\yng(1,1,1,1)}$ \\ \hline 
        Dimension & 1 & 3 & 2 & 3 & 1 \\ \hline 
    \end{tabular}
    \caption{Irreducible representations of $S_4$ and their dimensions, equal to
    the number of standard Young tableaux of each shape.}
    \label{tab:irrepdim}
\end{table}

\subsection{Semistandard Young tableaux and Schur--Weyl multiplicities}

A \emph{semistandard Young tableau} (SSYT) of shape $\lambda$ with
alphabet $\{1,\dots,d\}$ is a filling of the boxes of $\lambda$ with
entries from $\{1,\dots,d\}$ that is weakly increasing along rows
(left to right) and strictly increasing down columns.
The number of SSYTs of shape $\lambda$ with alphabet $\{1,\dots,d\}$
equals the dimension of the corresponding irreducible $\U(d)$
representation, and therefore gives the multiplicity with which the
$S_N$ irrep labeled by $\lambda$ appears in the Schur--Weyl
decomposition of $\HH_d^N$.
For a systematic treatment of this correspondence see, e.g.,
\cite{fultonharris}.

For $d=2$ (spin-$\tfrac{1}{2}$, alphabet $\{1,2\}$) and $N=4$, the
SSYTs of each admissible shape are:

\medskip
\noindent\text{Shape $(4)$:}\quad
${\tiny\Yvcentermath1\young(1111)}$,\quad
${\tiny\Yvcentermath1\young(1112)}$,\quad
${\tiny\Yvcentermath1\young(1122)}$,\quad
${\tiny\Yvcentermath1\young(1222)}$,\quad
${\tiny\Yvcentermath1\young(2222)}$
\quad---\quad multiplicity $\mathbf{5}$.

\medskip
\noindent\text{Shape $(3,1)$:}\quad
${\tiny\Yvcentermath1\young(111,2)}$,\quad
${\tiny\Yvcentermath1\young(112,2)}$,\quad
${\tiny\Yvcentermath1\young(122,2)}$
\quad---\quad multiplicity $\mathbf{3}$.

\medskip
\noindent\text{Shape $(2,2)$:}\quad
${\tiny\Yvcentermath1\young(11,22)}$
\quad---\quad multiplicity $\mathbf{1}$.

\medskip
These multiplicities $5,3,1$ are the prefactors appearing in
equation~\eqref{eq:spinhalf}.
Physically, they count the $2I+1$ magnetic substates of the
collective spin-$I$ multiplet:
\begin{align*}
    I=2 &\leftrightarrow\; 2(2)+1=5, \\
I=1 &\leftrightarrow\; 2(1)+1=3, \\
I=0 &\leftrightarrow\; 2(0)+1=1.
\end{align*}
More generally, for arbitrary single-nucleus spin $I$ (so $d=2I+1$),
the multiplicity of the $S_N$ irrep $\lambda$ in $\HH_d^N$ is given
by the number of SSYTs of shape $\lambda$ with alphabet
$\{1,\dots,d\}$, which equals the dimension of the corresponding
$\U(d)$ irrep and can be computed efficiently via the hook-content
formula \cite{fultonharris}.

\subsection{Descents and the major index}

Let $T$ be a standard Young tableau of shape $\lambda\vdash N$.  
A number $i\in\{1,\dots,N-1\}$ is a \emph{descent} of $T$ if the entry $i+1$ lies in a
strictly lower row than $i$.  
The \emph{major index} of $T$ is the sum of its descents:
\[
\mathrm{maj}(T)=\sum_{\substack{i=1 \\ i\text{ is a descent}}}^{N-1} i.
\]
For example,
\[
T = {\tiny\Yvcentermath1\young(12,3,4)}
\qquad\Rightarrow\qquad
\mathrm{maj}(T)=2+3=5.
\]

\subsection{Branching $S_N\rightarrow C_m$ via the major index}

The cyclic group $C_m=\langle r\rangle$ is generated by a single rotation $r$ satisfying
$r^m=1$.  
All of its irreducible representations are one-dimensional, given by
\[
\chi_j(r)=e^{2\pi i j/m},\qquad j=0,1,\dots,m-1.
\]

A theorem of Kra\'skiewicz--Weyman and Adin--Roichman
\cite{KraskiewiczWeyman1989,AdinRoichman2001} states that the restriction of an $S_N$
irrep to a cyclic subgroup is governed entirely by the major index of its standard
tableaux; the number of $\chi_j$
in the restricted representation is equal to the number of tableaux with major index congruent to $ j $ mod $ m $.
Equivalently, after restricting to $C_m$ one may choose a $C_m$--eigenbasis in which
the character $\chi_j$ occurs with exactly this multiplicity.  In particular, the
major index provides a purely combinatorial way to determine which cyclic
species appear, and how many times each species appears, without computing any
$S_N$ character values.

From the Schur--Weyl perspective, this branching rule converts the abstract
$S_N\times \U(d)$ decomposition into a practical tool for real molecules, whose
symmetry is almost always a proper subgroup of $S_N$.
It enables a direct passage from Young diagrams to the $C_m$ species relevant for
rovibrational spectroscopy, making the calculation of nuclear--spin statistical weights
transparent and automatable.

Unlike character tables, which decompose a representation only once it is already
known, the major-index rule furnishes a direct combinatorial passage from
the Schur--Weyl $S_N$-classification of the nuclear--spin space to the cyclic species
realized by the molecular geometry, uniformly across examples and higher-spin
isotopologues.

\section{Nuclear--Spin Species Under Cyclic Symmetry Reduction}

In realistic molecular settings the identical nuclei are constrained by a rigid or
quasi-rigid geometry, so the full permutation symmetry $S_N$ of $N$ identical nuclei is
reduced to the subgroup actually realized by the molecular symmetry group.  
Within the Schur--Weyl framework this means that, although
\[
\HH_d^N
\]
carries the canonical $S_N\times \U(d)$ action, only a subgroup of $S_N$ is physically
relevant.  For many polyatomic molecules this residual symmetry includes a
rotation about a principal molecular axis.  In this section we combine the Schur--Weyl
decomposition with the major--index branching rule to treat the cyclic restriction
\[
S_N \longrightarrow C_m
\]
and obtain nuclear--spin statistical weights directly from Young diagrams.

The examples below are organized by molecule for chemical concreteness, but the
underlying representation-theoretic structure is governed by the permutation subgroup
realized by the molecular geometry.  In particular, the examples illustrate the cases
of cyclic reduction $S_N\downarrow C_m$ for $m=3,4,6,7$, together with the recovery of
the corresponding dihedral or point-group species after adjoining reflections.

\subsection{Nuclear--spin species of \ce{XeOF4}}

\begin{figure}[htp]
    \centering
    \includegraphics[width=3cm]{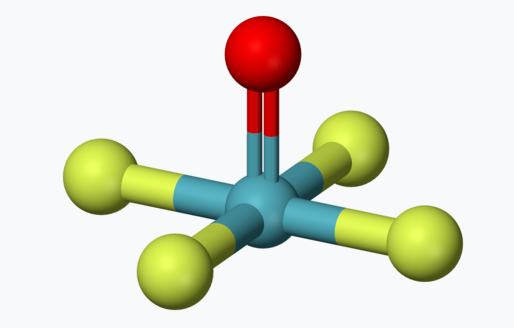}
    \caption{Xenon oxytetrafluoride}
    \label{fig:XeOF4}
\end{figure}

Xenon oxytetrafluoride \ce{XeOF4}, visualized in Figure~\ref{fig:XeOF4}, contains four identical fluorine nuclei, each
with spin $I=\tfrac12$.

Thus the fluorine spin Hilbert space decomposes through
Schur--Weyl duality as
\[
\HH_2^4
\;\cong\;
5\,{\tiny\Yvcentermath1\yng(4)}
\;\oplus\;
3\,{\tiny\Yvcentermath1\yng(3,1)}
\;\oplus\;
1\,{\tiny\Yvcentermath1\yng(2,2)}.
\]

The four fluorines lie at the corners of a square, and a $90^\circ$ rotation
about the principal axis permutes them by the $4$--cycle $(1\,2\,3\,4)$, so the
realized rotational permutation symmetry contains
\[
C_4=\langle(1\,2\,3\,4)\rangle\subset S_4.
\]
By the major--index branching rule, each standard Young tableau $T$ contributes
to the one--dimensional irrep $\chi_j$ of $C_4$ with
\[
\mathrm{maj}(T)\equiv j\pmod 4.
\]

\medskip\noindent
\text{Shape $(4)$:}
\begin{align*}
\mathrm{maj}\!\left({\tiny\Yvcentermath1 \young(1234)}\right)&\equiv 0.
\end{align*}

\noindent
\text{Shape $(3,1)$:}\qquad
\begin{align*}
\mathrm{maj}\!\left({\tiny\Yvcentermath1 \young(134,2)}\right)&\equiv 1, &
\mathrm{maj}\!\left({\tiny\Yvcentermath1 \young(124,3)}\right)&\equiv 2, &
\mathrm{maj}\!\left({\tiny\Yvcentermath1 \young(123,4)}\right)&\equiv 3.
\end{align*}

\noindent
\text{Shape $(2,2)$:}\qquad
\begin{align*}
\mathrm{maj}\!\left({\tiny\Yvcentermath1 \young(12,34)}\right)&\equiv 2, &
\mathrm{maj}\!\left({\tiny\Yvcentermath1 \young(13,24)}\right)&\equiv 0.
\end{align*}

Grouping by shape yields the restriction to $C_4$:
\[
\begin{aligned}
{\tiny\Yvcentermath1\yng(4)}   &\downarrow \chi_0,\\[3pt]
{\tiny\Yvcentermath1\yng(3,1)} &\downarrow \chi_1\oplus\chi_2\oplus\chi_3,\\[3pt]
{\tiny\Yvcentermath1\yng(2,2)} &\downarrow \chi_0\oplus\chi_2.
\end{aligned}
\]

To obtain point--group species one must include a reflection generator
to pass from $C_4$ to the point group $C_{4v}$.  Concretely, take the
standard reflection
\[
s=(2\,4),
\qquad\text{so}\qquad C_{4v}=\langle r,s\rangle,\quad r=(1\,2\,3\,4).
\]
The character table of $C_{4v}$ is given in
Table~\ref{tab:C4v}.

\begin{table}[h]
\centering
\begin{tabular}{c|ccccc}
$C_{4v}$  
& $E$ & $2C_4$ & $C_2$ & $2\sigma_v$ & $2\sigma_d$ \\ \hline
Permutation class
& $\{1\}$
& $\{r,r^3\}$
& $\{r^2\}$
& $\{s,r^2s\}$
& $\{rs,r^3s\}$ \\ \hline
$A_1$ &  1 &  1 &  1 &  1 &  1 \\
$A_2$ &  1 &  1 &  1 & $-1$ & $-1$ \\
$B_1$ &  1 & $-1$ &  1 &  1 & $-1$ \\
$B_2$ &  1 & $-1$ &  1 & $-1$ &  1 \\
$E$   &  2 &  0 & $-2$ &  0 &  0
\end{tabular}
\caption{Character table of $C_{4v}$ together with the induced permutation
classes on the four equivalent fluorine nuclei, where $r=(1234)$ and $s=(24)$.}
\label{tab:C4v}
\end{table}

The restriction of each $C_{4v}$ irrep to the rotational subgroup
$C_4=\langle r\rangle$ is determined by its character on $r$, since
$\chi_0(r)=1$, $\chi_2(r)=e^{2\pi i\cdot 2/4}=-1$,
and $(\chi_1\oplus\chi_3)(r)=i+(-i)=0$.  Reading off the $2C_4$
column of the character table therefore gives
\[
A_1\downarrow \chi_0,\quad
A_2\downarrow \chi_0,\quad
B_1\downarrow \chi_2,\quad
B_2\downarrow \chi_2,\quad
E\downarrow (\chi_1\oplus\chi_3).
\]
Thus $\chi_1\oplus\chi_3$ recombines uniquely as $E$, while $\chi_0$
and $\chi_2$ each split into two one--dimensional species distinguished
by the sign under the reflection $s$.

To determine this sign we use the Young symmetrizer rule: for the tableau
basis vectors obtained from Young symmetrizers, a transposition
$s=(a\;b)$ acts by $+1$ if $a$ and $b$ lie in the same row of
$T$ (row symmetrization) and by $-1$ if they lie in the same column of
$T$ (column antisymmetrization).  Irreps with eigenvalue $+1$ under $s$
carry subscript $1$ (i.e.\ $A_1$ or $B_1$), and those with eigenvalue
$-1$ carry subscript $2$ (i.e.\ $A_2$ or $B_2$).
Applying this to $s=(2\,4)$ on each $\chi_0$ and $\chi_2$ tableau:
\begin{align*}
{\tiny\Yvcentermath1\young(1234)}\,:&\quad
  2,4\in\text{row }1 \;\Rightarrow\; +1 \;\Rightarrow\; A_1,\\[4pt]
{\tiny\Yvcentermath1\young(13,24)}\,:&\quad
  2,4\in\text{row }2 \;\Rightarrow\; +1 \;\Rightarrow\; A_1,\\[4pt]
{\tiny\Yvcentermath1\young(124,3)}\,:&\quad
  2,4\in\text{row }1 \;\Rightarrow\; +1 \;\Rightarrow\; B_1,\\[4pt]
{\tiny\Yvcentermath1\young(12,34)}\,:&\quad
  2,4\in\text{col }2 \;\Rightarrow\; -1 \;\Rightarrow\; B_2.
\end{align*}

Thus we have the $S_4\downarrow C_{4v}$ branching
rules:
\[
\begin{aligned}
{\tiny\Yvcentermath1\yng(4)}   &\downarrow A_1,\\[3pt]
{\tiny\Yvcentermath1\yng(3,1)} &\downarrow B_1 \oplus E,\\[3pt]
{\tiny\Yvcentermath1\yng(2,2)} &\downarrow A_1 \oplus B_2.
\end{aligned}
\]

Combining these with the Schur--Weyl multiplicities $5,3,1$ we have that
$\HH_2^4$ branches down to $C_{4v}$ species as
\[
5A_1
\oplus
3(B_1\oplus E)
\oplus
(A_1\oplus B_2),
\]
yielding the total fluorine nuclear--spin representation for \ce{XeOF4}:
\[
\Gamma_{\mathrm{nspin}}^{\mathrm{tot}}
 \;\cong\;
6\,A_1
\;\oplus\;
0\,A_2
\;\oplus\;
3\,B_1
\;\oplus\;
1\,B_2
\;\oplus\;
3\,E,
\]
so the $C_{4v}$ nuclear--spin statistical weights are in the ratio
\[
A_{1}:A_{2}:B_{1}:B_{2}:E \;=\; 6:0:3:1:3,
\]
(with total dimension $6+0+3+1+2\cdot 3=16=2^4$ as required).

For completeness, one may also record the corresponding total-spin refinement.
In the spin-$\tfrac12$ cases this refinement is immediate from the two-row
Schur--Weyl decomposition.  Using
\[
{\tiny\Yvcentermath1\yng(4)} \leftrightarrow I=2,\qquad
{\tiny\Yvcentermath1\yng(3,1)} \leftrightarrow I=1,\qquad
{\tiny\Yvcentermath1\yng(2,2)} \leftrightarrow I=0,
\]
we obtain the refined contributions
\[
I=2:\ 5A_1,
\qquad
I=1:\ 3(B_1\oplus E),
\qquad
I=0:\ (A_1\oplus B_2).
\]

\subsection{Nuclear--spin species of benzene}

Benzene \ce{C6H6} contains six equivalent protons, each with spin $I=\tfrac12$.
Thus the proton spin Hilbert space decomposes through Schur--Weyl duality as
\[
\HH_2^6
\;\cong\;
7\,{\tiny\Yvcentermath1\yng(6)}
\;\oplus\;
5\,{\tiny\Yvcentermath1\yng(5,1)}
\;\oplus\;
3\,{\tiny\Yvcentermath1\yng(4,2)}
\;\oplus\;
1\,{\tiny\Yvcentermath1\yng(3,3)}.
\]

In rigid benzene, a $60^\circ$ rotation about the principal axis permutes the
protons by the $6$--cycle $(1\,2\,3\,4\,5\,6)$, so the realized rotational
permutation symmetry contains
\[
C_6=\langle(1\,2\,3\,4\,5\,6)\rangle\subset S_6.
\]
By the major--index branching rule, each standard Young tableau $T$ contributes
to the one--dimensional irrep $\chi_j$ of $C_6$ with
\[
\mathrm{maj}(T)\equiv j\pmod 6.
\]

\medskip\noindent
\text{Shape $(6)$:}
\begin{align*}
\mathrm{maj}\!\left({\tiny\Yvcentermath1 \young(123456)}\right)&\equiv 0.
\end{align*}

\noindent
\text{Shape $(5,1)$:}\qquad
\begin{align*}
\mathrm{maj}\!\left({\tiny\Yvcentermath1 \young(13456,2)}\right)&\equiv 1, &
\mathrm{maj}\!\left({\tiny\Yvcentermath1 \young(12456,3)}\right)&\equiv 2, &
\mathrm{maj}\!\left({\tiny\Yvcentermath1 \young(12356,4)}\right)&\equiv 3,\\[2pt]
\mathrm{maj}\!\left({\tiny\Yvcentermath1 \young(12346,5)}\right)&\equiv 4, &
\mathrm{maj}\!\left({\tiny\Yvcentermath1 \young(12345,6)}\right)&\equiv 5.
\end{align*}

\noindent
\text{Shape $(4,2)$:}\qquad
\begin{align*}
\mathrm{maj}\!\left({\tiny\Yvcentermath1 \young(1246,35)}\right)&\equiv 0, &
\mathrm{maj}\!\left({\tiny\Yvcentermath1 \young(1345,26)}\right)&\equiv 0,\\[2pt]
\mathrm{maj}\!\left({\tiny\Yvcentermath1 \young(1245,36)}\right)&\equiv 1,\\[2pt]
\mathrm{maj}\!\left({\tiny\Yvcentermath1 \young(1235,46)}\right)&\equiv 2, &
\mathrm{maj}\!\left({\tiny\Yvcentermath1 \young(1256,34)}\right)&\equiv 2,\\[2pt]
\mathrm{maj}\!\left({\tiny\Yvcentermath1 \young(1236,45)}\right)&\equiv 3,\\[2pt]
\mathrm{maj}\!\left({\tiny\Yvcentermath1 \young(1234,56)}\right)&\equiv 4, &
\mathrm{maj}\!\left({\tiny\Yvcentermath1 \young(1356,24)}\right)&\equiv 4,\\[2pt]
\mathrm{maj}\!\left({\tiny\Yvcentermath1 \young(1346,25)}\right)&\equiv 5.
\end{align*}

\noindent
\text{Shape $(3,3)$:}\qquad
\begin{align*}
\mathrm{maj}\!\left({\tiny\Yvcentermath1 \young(124,356)}\right)&\equiv 0, &
\mathrm{maj}\!\left({\tiny\Yvcentermath1 \young(125,346)}\right)&\equiv 1,\\[2pt]
\mathrm{maj}\!\left({\tiny\Yvcentermath1 \young(123,456)}\right)&\equiv 3, &
\mathrm{maj}\!\left({\tiny\Yvcentermath1 \young(135,246)}\right)&\equiv 3,\\[2pt]
\mathrm{maj}\!\left({\tiny\Yvcentermath1 \young(134,256)}\right)&\equiv 5.
\end{align*}

Grouping by shape yields the restriction to $C_6$:
\[
\begin{aligned}
{\tiny\Yvcentermath1\yng(6)}   &\downarrow \chi_0,\\[3pt]
{\tiny\Yvcentermath1\yng(5,1)} &\downarrow \chi_1\oplus\chi_2\oplus\chi_3\oplus\chi_4\oplus\chi_5,\\[3pt]
{\tiny\Yvcentermath1\yng(4,2)} &\downarrow 2\chi_0\oplus\chi_1\oplus 2\chi_2\oplus\chi_3\oplus 2\chi_4\oplus\chi_5,\\[3pt]
{\tiny\Yvcentermath1\yng(3,3)} &\downarrow \chi_0\oplus\chi_1\oplus 2\chi_3\oplus\chi_5.
\end{aligned}
\]

To obtain point--group species one must include a reflection generator to pass
from $C_6$ to the point group $D_{6h}$.  Concretely, take the standard dihedral
reflection
\[
s=(2\,6)(3\,5),
\qquad\text{so}\qquad D_6=\langle r,s\rangle\subset S_6,\quad r=(1\,2\,3\,4\,5\,6).
\]
The irreps of $D_6$ restrict to $C_6$ as
\begin{align*}
   & A_1\downarrow\chi_0,\quad
A_2\downarrow\chi_0,\quad \\
&B_1\downarrow\chi_3,\quad
B_2\downarrow\chi_3,\quad \\
& E_1\downarrow(\chi_1\oplus\chi_5),\quad
E_2\downarrow(\chi_2\oplus\chi_4) 
\end{align*}

Thus $\chi_1\oplus\chi_5$ and $\chi_2\oplus\chi_4$ recombine uniquely as $E_1$
and $E_2$, while $\chi_0$ and $\chi_3$ each split into two one--dimensional
species depending on the sign under the reflection $s$.  Evaluating $s$ on the $ \chi_0 $ tableaux to find the proper subscript for $ A $ and evaluating $ s $ on the $ \chi_3 $ tableaux to find the proper subscript for $ B $ gives the $S_6\downarrow D_6$ branching rules:
\[
\begin{aligned}
{\tiny\Yvcentermath1\yng(6)}   &\downarrow A_1,\\[3pt]
{\tiny\Yvcentermath1\yng(5,1)} &\downarrow B_1 \oplus E_1 \oplus E_2,\\[3pt]
{\tiny\Yvcentermath1\yng(4,2)} &\downarrow 2A_1 \oplus B_2 \oplus E_1 \oplus 2E_2,\\[3pt]
{\tiny\Yvcentermath1\yng(3,3)} &\downarrow A_2 \oplus 2B_1 \oplus E_1.
\end{aligned}
\]
Combining these with the Schur--Weyl multiplicities $7,5,3,1$ we have that
$\HH_2^6$ branches down to $D_6$ species as
\[
7A_1
\oplus
5(B_1 \oplus E_1 \oplus E_2)
\oplus
3(2A_1\oplus B_2 \oplus E_1 \oplus 2E_2)
\oplus
(A_2 \oplus 2B_1 \oplus E_1)
\]
yielding the total proton nuclear--spin representation for benzene:
\[
\Gamma_{\mathrm{nspin}}^{\mathrm{tot}}
 \;\cong\;
13\,A_1
\;\oplus\;
1\,A_2
\;\oplus\;
7\,B_1
\;\oplus\;
3\,B_2
\;\oplus\;
9\,E_1
\;\oplus\;
11\,E_2,
\]
so the $D_{6h}$ nuclear--spin statistical weights are in the ratio
\[
A_{1g}:A_{2g}:B_{1g}:B_{2g}:E_{1g}:E_{2g} \;=\; 13:1:7:3:9:11,
\]
(with total dimension $13+1+7+3+2(9+11)=64=2^6$ as required).

For completeness, one may also record the corresponding total-spin refinement.
In the spin-$\tfrac12$ cases this refinement is immediate from the two-row
Schur--Weyl decomposition. Using
\[
(6)\leftrightarrow I=3,
(5,1)\leftrightarrow I=2,
(4,2)\leftrightarrow I=1,
(3,3)\leftrightarrow I=0,
\]
we obtain the refined contributions given in Table~\ref{tab:C6H6}.
\label{tab:C6H6}

\begin{table}[ht]
\centering
\caption{Spin--resolved nuclear--spin species for benzene \ce{C6H6}.}
\begin{tabular}{c c}
\hline
 $I$ &  Species \\ 
\hline
$3$ & \;$7A_1$ \\[2pt]
$2$ & \;$5B_1 \oplus 5E_1 \oplus 5E_2$ \\[2pt]
$1$ & \; $6A_1 \oplus 3B_2 \oplus 3E_1 \oplus 6E_2$ \\[2pt]
$0$ & \;$A_2 \oplus 2B_1 \oplus E_1$ \\
\hline
\end{tabular}
\label{tab:C6H6}
\end{table}

\subsection{Nuclear--spin species of deuterated benzene \ce{C6D6}}

Deuterated benzene \ce{C6D6} contains six equivalent deuterium
nuclei, each with spin $I=1$, so 
\[
d=2I+1=3.
\]
For the spin-$\tfrac12$ examples above, the correspondence between two-row Young
shapes and total spin is immediate; the deuterated benzene example shows that this
simplification is special to $d=2$ and does not persist for higher-spin nuclei.

This example extends the benzene analysis of the previous section to $d=3$,
illustrating how the framework accommodates higher nuclear spin without
modification to the combinatorial procedure. 

This comparison isolates the effect of the single-nucleus spin: the realized
permutation subgroup and point-group reduction are unchanged from ordinary benzene,
but increasing the nuclear spin from $I=\tfrac12$ to $I=1$ changes the Schur--Weyl
decomposition by allowing Young diagrams with three rows (so the Schur--Weyl multiplicities are now counted by semistandard Young
tableaux with alphabet $\{1,2,3\}$), which in turn alters the
resulting statistical weights.

The geometric symmetry and permutation subgroup are the same as for
ordinary benzene: the deuterons are permuted by the $6$--cycle
$(1\,2\,3\,4\,5\,6)$, so the same cyclic reduction $ S_6\downarrow C_6
 $ and the same dihedral extension to $D_6$ apply.  Accordingly, the
major--index data for the shapes
\[
{\tiny\Yvcentermath1\yng(6)}
 \qquad
{\tiny\Yvcentermath1\yng(5,1)}
 \qquad
{\tiny\Yvcentermath1\yng(4,2)}
 \qquad
{\tiny\Yvcentermath1\yng(3,3)}
\]
are exactly the same as in the proton case and will not be repeated here.
The only new ingredients are the three--row shapes
\[
{\tiny\Yvcentermath1\yng(4,1,1)}
 \qquad
{\tiny\Yvcentermath1\yng(3,2,1)}
 \qquad
{\tiny\Yvcentermath1\yng(2,2,2)}
\]
which are absent for $d=2$.

The deuterium spin Hilbert space $ \HH_3^6 $ decomposes through Schur--Weyl
duality as
\begin{align*}
    &28\,{\tiny\Yvcentermath1\yng(6)}
\oplus \;
35\,{\tiny\Yvcentermath1\yng(5,1)}
\oplus \;
27\,{\tiny\Yvcentermath1\yng(4,2)}
\\
&\oplus \;
10\,{\tiny\Yvcentermath1\yng(3,3)}
\oplus \;
10\,{\tiny\Yvcentermath1\yng(4,1,1)}
\oplus \;
8\,{\tiny\Yvcentermath1\yng(3,2,1)}
\oplus \;
1\,{\tiny\Yvcentermath1\yng(2,2,2)}
\end{align*}
The multiplicities $28,35,27,10,10,8,1$ are the dimensions of the
corresponding $\U(3)$ irreps, obtained by counting semistandard Young
tableaux with alphabet $\{1,2,3\}$.  The total dimension is
$28\cdot 1+35\cdot 5+27\cdot 9+10\cdot 5+10\cdot 10+8\cdot 16+1\cdot 5
=729=3^6$, as required.  

For the four shapes already present in \ce{C6H6}, the restriction
to $C_6$ is unchanged.  We therefore record only the major--index
data for the new three--row shapes.

\noindent
\text{Shape $(4,1,1)$:}\qquad
\begin{align*}
\mathrm{maj}\!\left({\tiny\Yvcentermath1\young(1246,3,5)}\right)&\equiv 0, &
\mathrm{maj}\!\left({\tiny\Yvcentermath1\young(1345,2,6)}\right)&\equiv 0,\\[2pt]
\mathrm{maj}\!\left({\tiny\Yvcentermath1\young(1236,4,5)}\right)&\equiv 1, &
\mathrm{maj}\!\left({\tiny\Yvcentermath1\young(1245,3,6)}\right)&\equiv 1,\\[2pt]
\mathrm{maj}\!\left({\tiny\Yvcentermath1\young(1235,4,6)}\right)&\equiv 2,\\[2pt]
\mathrm{maj}\!\left({\tiny\Yvcentermath1\young(1234,5,6)}\right)&\equiv 3, &
\mathrm{maj}\!\left({\tiny\Yvcentermath1\young(1456,2,3)}\right)&\equiv 3,\\[2pt]
\mathrm{maj}\!\left({\tiny\Yvcentermath1\young(1356,2,4)}\right)&\equiv 4,\\[2pt]
\mathrm{maj}\!\left({\tiny\Yvcentermath1\young(1256,3,4)}\right)&\equiv 5, &
\mathrm{maj}\!\left({\tiny\Yvcentermath1\young(1346,2,5)}\right)&\equiv 5.
\end{align*}

\noindent
\text{Shape $(3,2,1)$:}\qquad
\begin{align*}
\mathrm{maj}\!\left({\tiny\Yvcentermath1\young(124,36,5)}\right)&\equiv 0, &
\mathrm{maj}\!\left({\tiny\Yvcentermath1\young(126,34,5)}\right)&\equiv 0,\\[2pt]
\mathrm{maj}\!\left({\tiny\Yvcentermath1\young(123,46,5)}\right)&\equiv 1, &
\mathrm{maj}\!\left({\tiny\Yvcentermath1\young(125,34,6)}\right)&\equiv 1, &
\mathrm{maj}\!\left({\tiny\Yvcentermath1\young(146,25,3)}\right)&\equiv 1,\\[2pt]
\mathrm{maj}\!\left({\tiny\Yvcentermath1\young(123,45,6)}\right)&\equiv 2, &
\mathrm{maj}\!\left({\tiny\Yvcentermath1\young(136,24,5)}\right)&\equiv 2, &
\mathrm{maj}\!\left({\tiny\Yvcentermath1\young(145,26,3)}\right)&\equiv 2,\\[2pt]
\mathrm{maj}\!\left({\tiny\Yvcentermath1\young(135,24,6)}\right)&\equiv 3, &
\mathrm{maj}\!\left({\tiny\Yvcentermath1\young(135,26,4)}\right)&\equiv 3,\\[2pt]
\mathrm{maj}\!\left({\tiny\Yvcentermath1\young(125,36,4)}\right)&\equiv 4, &
\mathrm{maj}\!\left({\tiny\Yvcentermath1\young(134,25,6)}\right)&\equiv 4, &
\mathrm{maj}\!\left({\tiny\Yvcentermath1\young(136,25,4)}\right)&\equiv 4,\\[2pt]
\mathrm{maj}\!\left({\tiny\Yvcentermath1\young(124,35,6)}\right)&\equiv 5, &
\mathrm{maj}\!\left({\tiny\Yvcentermath1\young(126,35,4)}\right)&\equiv 5, &  
\mathrm{maj}\!\left({\tiny\Yvcentermath1\young(134,26,5)}\right)&\equiv 5.
\end{align*}

\noindent
\text{Shape $(2,2,2)$:}\qquad
\begin{align*}
\mathrm{maj}\!\left({\tiny\Yvcentermath1\young(12,34,56)}\right)&\equiv 0, &
\mathrm{maj}\!\left({\tiny\Yvcentermath1\young(14,25,36)}\right)&\equiv 0,\\[2pt]
\mathrm{maj}\!\left({\tiny\Yvcentermath1\young(13,24,56)}\right)&\equiv 2,\\[2pt]
\mathrm{maj}\!\left({\tiny\Yvcentermath1\young(13,25,46)}\right)&\equiv 3,\\[2pt]
\mathrm{maj}\!\left({\tiny\Yvcentermath1\young(12,35,46)}\right)&\equiv 4.
\end{align*}

Grouping these tableaux by major index gives the additional
$C_6$-branching rules
\[
\begin{aligned}
{\tiny\Yvcentermath1\yng(4,1,1)} &\downarrow
2\chi_0\oplus 2\chi_1\oplus\chi_2\oplus 2\chi_3\oplus\chi_4\oplus 2\chi_5,\\[3pt]
{\tiny\Yvcentermath1\yng(3,2,1)} &\downarrow
2\chi_0\oplus 3\chi_1\oplus 3\chi_2\oplus 2\chi_3\oplus 3\chi_4\oplus 3\chi_5,\\[3pt]
{\tiny\Yvcentermath1\yng(2,2,2)} &\downarrow
2\chi_0\oplus\chi_2\oplus\chi_3\oplus\chi_4.
\end{aligned}
\]
Together with the proton-case branchings for
\[
(6),\ (5,1),\ (4,2),\ (3,3),
\]
this gives the full restriction $S_6\downarrow C_6$ for deuterated benzene. The passage from $C_6$ to $D_6$ is again the same as in the benzene
case.  For the four two--row shapes, the $S_6\downarrow D_6$ branching
is therefore unchanged.  For the new three--row shapes, the branching is again obtained by evaluating $s$ on the $ \chi_0 $ tableaux to find the proper subscript for $ A $ and evaluating $ s $ on the $ \chi_3 $ tableaux to find the proper subscript for $ B $, yielding the additional $S_6\downarrow D_6$ branching rules
\[
\begin{aligned}
{\tiny\Yvcentermath1\yng(4,1,1)} &\downarrow 2A_2\oplus B_1\oplus B_2\oplus 2E_1\oplus E_2,\\[3pt]
{\tiny\Yvcentermath1\yng(3,2,1)} &\downarrow A_1\oplus A_2\oplus B_1\oplus B_2\oplus 3E_1\oplus 3E_2,\\[3pt]
{\tiny\Yvcentermath1\yng(2,2,2)} &\downarrow 2A_1\oplus B_2\oplus E_2.
\end{aligned}
\]
Combining these with the Schur--Weyl multiplicities $28,35,27,10,10,8,1$
we have that $\HH_3^6$ branches down to $D_6$ species as
\begin{align*}
&28\,A_1\\
\oplus\;&35\,(B_1\oplus E_1\oplus E_2)\\
\oplus\;&27\,(2A_1\oplus B_2\oplus E_1\oplus 2E_2)\\
\oplus\;&10\,(A_2\oplus 2B_1\oplus E_1)\\
\oplus\;&10\,(2A_2\oplus B_1\oplus B_2\oplus 2E_1\oplus E_2)\\
\oplus\;&8\,(A_1\oplus A_2\oplus B_1\oplus B_2\oplus 3E_1\oplus 3E_2)\\
\oplus\;&1\,(2A_1\oplus B_2\oplus E_2),
\end{align*}
yielding the total deuterium nuclear--spin representation for
\ce{C6D6}:
\begin{align*}
\Gamma_{\mathrm{nspin}}^{\mathrm{tot}}
&\;\cong\;
92\,A_1
\;\oplus\;
38\,A_2
\;\oplus\;
73\,B_1
 \\
 &\;\oplus\;
46\,B_2
\;\oplus\;
116\,E_1
\;\oplus\;
124\,E_2,
\end{align*}
so the $D_{6h}$ nuclear--spin statistical weights are in the ratio
\[
A_1:A_2:B_1:B_2:E_1:E_2 \;=\; 92:38:73:46:116:124,
\]
with total dimension
$92+38+73+46+2(116+124)=729=3^6$ as required.

Comparing with the proton case \ce{C6H6} (Section~III.C), the
statistical weights shift from $13:1:7:3:9:11$ to $92:38:73:46:116:124$.
The structural difference arises from the additional three--row
shapes $(4,1,1)$, $(3,2,1)$, and $(2,2,2)$, which are absent in the
$d=2$ decomposition but contribute substantially here.

We can refine the \ce{C6D6} analysis further by branching the relevant
$\SU3$ multiplicity spaces to the physical spin subgroup
\[
\SU2 \hookrightarrow \SU3,
\]
for which the defining $3$-dimensional representation of $\SU3$
restricts to the spin-$1$ irrep of $\SU2$. Unlike the proton case, the Young shape no longer determines a unique total-spin
sector for $d=3$; the higher-spin setting therefore exhibits genuinely new structure
beyond the two-row correspondence $I=(\lambda_1-\lambda_2)/2$.
Indeed, under this embedding one finds
\[
\begin{aligned}
{\tiny\Yvcentermath1\yng(6)} &\downarrow (I=6) \oplus (I=4) \oplus (I=2 )\oplus (I=0),\\[3pt]
{\tiny\Yvcentermath1\yng(5,1)} &\downarrow (I=5) \oplus (I=4) \oplus (I=3) \oplus (I=2) \oplus (I=1),\\[3pt]
{\tiny\Yvcentermath1\yng(4,2)} &\downarrow (I=4) \oplus (I=3) \oplus 2(I=2) \oplus (I=0),\\[3pt]
{\tiny\Yvcentermath1\yng(4,1,1)} &\downarrow (I=3) \oplus (I=1),\\[3pt]
{\tiny\Yvcentermath1\yng(3,3)} &\downarrow (I=3) \oplus (I=1),\\[3pt]
{\tiny\Yvcentermath1\yng(3,2,1)} &\downarrow (I=2) \oplus (I=1),\\[3pt]
{\tiny\Yvcentermath1\yng(2,2,2)} &\downarrow (I=0).
\end{aligned}
\]
Combining this with the $S_6\downarrow D_6$ branching rules gives the
spin--resolved nuclear--spin species for \ce{C6D6} given in Table~\ref{tab:C6D6}.

\begin{table}[ht]
\centering
\caption{Spin--resolved nuclear--spin species for deuterated benzene \ce{C6D6}.}
\begin{tabular}{c c}
\hline
$I$ & Species \\
\hline
$6$ & $13\,A_1$ \\[2pt]
$5$ & $11\,B_1 \oplus 11\,E_1 \oplus 11\,E_2$ \\[2pt]
$4$ & $27\,A_1 \oplus 9\,B_1 \oplus 9\,B_2 \oplus 18\,E_1 \oplus 27\,E_2$ \\[2pt]
$3$ & $14\,A_1 \oplus 21\,A_2 \oplus 28\,B_1 \oplus 14\,B_2 \oplus 35\,E_1 \oplus 28\,E_2$ \\[2pt]
$2$ & $30\,A_1 \oplus 5\,A_2 \oplus 10\,B_1 \oplus 15\,B_2 \oplus 30\,E_1 \oplus 40\,E_2$ \\[2pt]
$1$ & $3\,A_1 \oplus 12\,A_2 \oplus 15\,B_1 \oplus 6\,B_2 \oplus 21\,E_1 \oplus 15\,E_2$ \\[2pt]
$0$ & $5\,A_1 \oplus 2\,B_2 \oplus E_1 \oplus 3\,E_2$ \\
\hline
\end{tabular}
\label{tab:C6D6}
\end{table}

Summing over $I$ recovers
\[
\Gamma_{\mathrm{nspin}}^{\mathrm{tot}}
\cong
92\,A_1
\oplus
38\,A_2
\oplus
73\,B_1
\oplus
46\,B_2
\oplus
116\,E_1
\oplus
124\,E_2.
\]

\subsection{Nuclear--spin species of  tropylium cation}

The tropylium cation \ce{C7H7+} contains seven equivalent protons, each with spin
$I=\tfrac12$. Thus the proton spin Hilbert space decomposes through
Schur--Weyl duality as
\[
\HH_2^7
\;\cong\;
8\,{\tiny\Yvcentermath1\yng(7)}
\;\oplus\;
6\,{\tiny\Yvcentermath1\yng(6,1)}
\;\oplus\;
4\,{\tiny\Yvcentermath1\yng(5,2)}
\;\oplus\;
2\,{\tiny\Yvcentermath1\yng(4,3)}.
\]

In rigid tropylium, a $2\pi/7$ rotation about the principal axis permutes the
protons by the $7$--cycle $(1\,2\,3\,4\,5\,6\,7)$, so the realized rotational
permutation symmetry contains
\[
C_7=\langle(1\,2\,3\,4\,5\,6\,7)\rangle\subset S_7.
\]
By the major--index branching rule, each standard Young tableau $T$ contributes
to the one--dimensional irrep $\chi_j$ of $C_7$ with
\[
\mathrm{maj}(T)\equiv j\pmod 7.
\]

\medskip\noindent
\text{Shape $(7)$:}
\begin{align*}
\mathrm{maj}\!\left({\tiny\Yvcentermath1 \young(1234567)}\right)&\equiv 0.
\end{align*}

\noindent
\text{Shape $(6,1)$:}\qquad
\begin{align*}
\mathrm{maj}\!\left({\tiny\Yvcentermath1 \young(134567,2)}\right)&\equiv 1, &
\mathrm{maj}\!\left({\tiny\Yvcentermath1 \young(124567,3)}\right)&\equiv 2,\\
\mathrm{maj}\!\left({\tiny\Yvcentermath1 \young(123567,4)}\right)&\equiv 3,&
\mathrm{maj}\!\left({\tiny\Yvcentermath1 \young(123467,5)}\right)&\equiv 4, \\
\mathrm{maj}\!\left({\tiny\Yvcentermath1 \young(123457,6)}\right)&\equiv 5, &
\mathrm{maj}\!\left({\tiny\Yvcentermath1 \young(123456,7)}\right)&\equiv 6.
\end{align*}

\noindent
\text{Shape $(5,2)$:}\qquad
\begin{align*}
\mathrm{maj}\!\left({\tiny\Yvcentermath1 \young(12457,36)}\right)&\equiv 0, &
\mathrm{maj}\!\left({\tiny\Yvcentermath1 \young(13456,27)}\right)&\equiv 0,\\[2pt]
\mathrm{maj}\!\left({\tiny\Yvcentermath1 \young(12357,46)}\right)&\equiv 1, &
\mathrm{maj}\!\left({\tiny\Yvcentermath1 \young(12456,37)}\right)&\equiv 1,\\[2pt]
\mathrm{maj}\!\left({\tiny\Yvcentermath1 \young(12567,34)}\right)&\equiv 2, &
\mathrm{maj}\!\left({\tiny\Yvcentermath1 \young(12356,47)}\right)&\equiv 2,\\[2pt]
\mathrm{maj}\!\left({\tiny\Yvcentermath1 \young(12367,45)}\right)&\equiv 3, &
\mathrm{maj}\!\left({\tiny\Yvcentermath1 \young(12346,57)}\right)&\equiv 3,\\[2pt]
\mathrm{maj}\!\left({\tiny\Yvcentermath1 \young(12347,56)}\right)&\equiv 4, &
\mathrm{maj}\!\left({\tiny\Yvcentermath1 \young(13567,24)}\right)&\equiv 4,\\[2pt]
\mathrm{maj}\!\left({\tiny\Yvcentermath1 \young(12345,67)}\right)&\equiv 5, &
\mathrm{maj}\!\left({\tiny\Yvcentermath1 \young(13467,25)}\right)&\equiv 5,\\[2pt]
\mathrm{maj}\!\left({\tiny\Yvcentermath1 \young(12467,35)}\right)&\equiv 6, &
\mathrm{maj}\!\left({\tiny\Yvcentermath1 \young(13457,26)}\right)&\equiv 6.
\end{align*}

\noindent
\text{Shape $(4,3)$:}\qquad
\begin{align*}
\mathrm{maj}\!\left({\tiny\Yvcentermath1 \young(1245,367)}\right)&\equiv 0, &
\mathrm{maj}\!\left({\tiny\Yvcentermath1 \young(1257,346)}\right)&\equiv 0,\\[2pt]
\mathrm{maj}\!\left({\tiny\Yvcentermath1 \young(1235,467)}\right)&\equiv 1, &
\mathrm{maj}\!\left({\tiny\Yvcentermath1 \young(1256,347)}\right)&\equiv 1,\\[2pt]
\mathrm{maj}\!\left({\tiny\Yvcentermath1 \young(1236,457)}\right)&\equiv 2, &
\mathrm{maj}\!\left({\tiny\Yvcentermath1 \young(1357,246)}\right)&\equiv 2,\\[2pt]
\mathrm{maj}\!\left({\tiny\Yvcentermath1 \young(1237,456)}\right)&\equiv 3, &
\mathrm{maj}\!\left({\tiny\Yvcentermath1 \young(1356,247)}\right)&\equiv 3,\\[2pt]
\mathrm{maj}\!\left({\tiny\Yvcentermath1 \young(1234,567)}\right)&\equiv 4, &
\mathrm{maj}\!\left({\tiny\Yvcentermath1 \young(1346,257)}\right)&\equiv 4,\\[2pt]
\mathrm{maj}\!\left({\tiny\Yvcentermath1 \young(1347,256)}\right)&\equiv 5, &
\mathrm{maj}\!\left({\tiny\Yvcentermath1 \young(1246,357)}\right)&\equiv 5,\\[2pt]
\mathrm{maj}\!\left({\tiny\Yvcentermath1 \young(1247,356)}\right)&\equiv 6, &
\mathrm{maj}\!\left({\tiny\Yvcentermath1 \young(1345,267)}\right)&\equiv 6.
\end{align*}

Grouping by shape yields the restriction to $C_7$:
\[
\begin{aligned}
{\tiny\Yvcentermath1\yng(7)}   &\downarrow \chi_0,\\[3pt]
{\tiny\Yvcentermath1\yng(6,1)} &\downarrow \chi_1\oplus\chi_2\oplus\chi_3\oplus\chi_4\oplus\chi_5\oplus\chi_6,\\[3pt]
{\tiny\Yvcentermath1\yng(5,2)} &\downarrow 2\chi_0\oplus 2\chi_1\oplus 2\chi_2\oplus 2\chi_3\oplus 2\chi_4\oplus 2\chi_5\oplus 2\chi_6,\\[3pt]
{\tiny\Yvcentermath1\yng(4,3)} &\downarrow 2\chi_0\oplus 2\chi_1\oplus 2\chi_2\oplus 2\chi_3\oplus 2\chi_4\oplus 2\chi_5\oplus 2\chi_6.
\end{aligned}
\]

To obtain point--group species one must include a reflection generator to pass
from $C_7$ to the point group $D_{7h}$.  Concretely, take the standard dihedral
reflection
\[
s=(2\,7)(3\,6)(4\,5),
\qquad\text{so}\qquad D_7=\langle r,s\rangle\subset S_7.
\]
For odd $n=7$, the dihedral group has two one--dimensional irreps $A_1,A_2$ (both
restricting to $\chi_0$ on $C_7$) and three two--dimensional irreps $E_1,E_2,E_3$
(corresponding to $\chi_j\oplus\chi_{7-j}$ for $j=1,2,3$).  The resulting $D_7$
branching of the relevant Specht irreps is
\[
\begin{aligned}
{\tiny\Yvcentermath1\yng(7)}   &\downarrow A_1,\\[3pt]
{\tiny\Yvcentermath1\yng(6,1)} &\downarrow E_1 \oplus E_2 \oplus E_3,\\[3pt]
{\tiny\Yvcentermath1\yng(5,2)} &\downarrow 2A_1 \oplus 2E_1 \oplus 2E_2 \oplus 2E_3,\\[3pt]
{\tiny\Yvcentermath1\yng(4,3)} &\downarrow A_1 \oplus A_2 \oplus 2E_1 \oplus 2E_2 \oplus 2E_3.
\end{aligned}
\]


(As in benzene, the physical point group is $D_{7h}$; for the nuclear--spin
permutation representation one may append the appropriate $g/u$ label uniformly,
so we suppress it here.)

Combining these with the Schur--Weyl multiplicities $8,6,4,2$ we have the total proton nuclear--spin representation for tropylium:
\[
\Gamma_{\mathrm{nspin}}^{\mathrm{tot}}
 \;\cong\;
18\,A_1
\;\oplus\;
2\,A_2
\;\oplus\;
18\,E_1
\;\oplus\;
18\,E_2
\;\oplus\;
18\,E_3,
\]
so the $D_{7h}$ nuclear--spin statistical weights are in the ratio
\[
A_{1}:A_{2}:E_{1}:E_{2}:E_{3} \;=\; 18:2:18:18:18,
\]
(with total dimension $18+2+2(18+18+18)=128=2^7$ as required).

For completeness, one may also record the corresponding total-spin refinement.
In the spin-$\tfrac12$ cases this refinement is immediate from the two-row
Schur--Weyl decomposition.  Using
\[
(7)\leftrightarrow I=\tfrac72,
(6,1)\leftrightarrow I=\tfrac52,
(5,2)\leftrightarrow I=\tfrac32,
(4,3)\leftrightarrow I=\tfrac12
\]
we obtain the refined contributions given in Table~\ref{tab:C7H7}.

\begin{table}[ht]
\centering
\caption{Spin--resolved nuclear--spin species for the tropylium cation \ce{C7H7+}.}
\begin{tabular}{c c}
\hline
 $I$ & Species \\ 
\hline
$\tfrac{7}{2}$ & $8\,A_1$ \\[2pt]
$\tfrac{5}{2}$ & $6E_1 \oplus 6E_2 \oplus 6E_3$ \\[2pt]
$\tfrac{3}{2}$ & $8A_1 \oplus 8E_1 \oplus 8E_2 \oplus 8E_3$ \\[2pt]
$\tfrac{1}{2}$ & $2A_1 \oplus 2A_2 \oplus 4E_1 \oplus 4E_2 \oplus 4E_3$ \\
\hline
\end{tabular}
\label{tab:C7H7}
\end{table}

\subsection{Nuclear--spin species of \ce{SF4}}

\begin{figure}[htp]
    \centering
    \includegraphics[width=3cm]{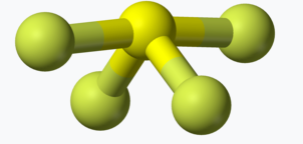}
    \caption{Sulfur tetrafluoride}
    \label{fig:SF4}
\end{figure}

Sulfur tetrafluoride \ce{SF4}, visualized in Figure~\ref{fig:SF4}, contains four equivalent fluorine nuclei, each with
spin $I=\tfrac12$. 

The fluorine spin Hilbert space therefore decomposes through
Schur--Weyl duality as
\[
\HH_2^4
\;\cong\;
5\,{\tiny\Yvcentermath1\yng(4)}
\;\oplus\;
3\,{\tiny\Yvcentermath1\yng(3,1)}
\;\oplus\;
1\,{\tiny\Yvcentermath1\yng(2,2)}.
\]

In contrast to benzene or tropylium, the cyclic symmetry of \ce{SF4} does not act
as a single $N$--cycle on the identical nuclei.  In the rigid molecule there is a
$C_2$ rotation that exchanges the two axial fluorines and simultaneously exchanges
the two equatorial fluorines.  Labeling the axial fluorines by $1,2$ and the
equatorial fluorines by $3,4$, the induced action on the nuclei is
\[
C_2 \hookrightarrow S_4,
\qquad
r \mapsto (1\,2)(3\,4).
\]
More generally, if a generator of $C_m$ embeds into $S_N$ as a product of disjoint
$m$--cycles, the major--index statistic must be taken modulo $m$.  For \ce{SF4},
this means that $\mathrm{maj}(T)$ is reduced modulo $2$.

By the major--index branching rule, each standard Young tableau $T$ contributes to
the one--dimensional irrep $\chi_j$ of $C_2$ with
\[
\mathrm{maj}(T)\equiv j\pmod 2,
\]
where $\chi_0$ and $\chi_1$ are the trivial and sign characters of $C_2$,
respectively.

\medskip\noindent
\text{Shape $(4)$:}
\begin{align*}
\mathrm{maj}\!\left({\tiny\Yvcentermath1 \young(1234)}\right)&\equiv 0.
\end{align*}

\noindent
\text{Shape $(3,1)$:}\qquad
\begin{align*}
\mathrm{maj}\!\left({\tiny\Yvcentermath1 \young(134,2)}\right)&\equiv 1, &
\mathrm{maj}\!\left({\tiny\Yvcentermath1 \young(124,3)}\right)&\equiv 0, &
\mathrm{maj}\!\left({\tiny\Yvcentermath1 \young(123,4)}\right)&\equiv 1.
\end{align*}

\noindent
\text{Shape $(2,2)$:}\qquad
\begin{align*}
\mathrm{maj}\!\left({\tiny\Yvcentermath1 \young(12,34)}\right)&\equiv 0, &
\mathrm{maj}\!\left({\tiny\Yvcentermath1 \young(13,24)}\right)&\equiv 0.
\end{align*}

Grouping by Young shape yields the restriction to $C_2$:
\[
\begin{aligned}
{\tiny\Yvcentermath1\yng(4)}        &\downarrow \chi_0,\\[3pt]
{\tiny\Yvcentermath1\yng(3,1)}      &\downarrow \chi_0 \oplus 2\chi_1,\\[3pt]
{\tiny\Yvcentermath1\yng(2,2)}      &\downarrow 2\chi_0,\\[3pt]
\end{aligned}
\]

To obtain point--group species one must include a reflection generator to pass
from $C_2$ to the point group $C_{2v}$.  A convenient choice is the vertical
reflection
\[
s=(1\,2),
\qquad\text{so}\qquad C_{2v}=\langle r,s\rangle.
\]
The one--dimensional irreps of $C_{2v}$ are distinguished by their characters on
$r$ and $s$, and the sign under $s$ can be determined directly from the Young
symmetrizer: tableaux with $1$ and $2$ in the same row are symmetric under $s$,
while those with $1$ and $2$ in the same column are antisymmetric.

Carrying this out yields the $S_4\downarrow C_{2v}$ branching rules:
\[
\begin{aligned}
{\tiny\Yvcentermath1\yng(4)}        &\downarrow A_1,\\[3pt]
{\tiny\Yvcentermath1\yng(3,1)}      &\downarrow A_1 \oplus B_1 \oplus B_2,\\[3pt]
{\tiny\Yvcentermath1\yng(2,2)}      &\downarrow A_1 \oplus A_2,\\[3pt]
\end{aligned}
\]

Combining these with the Schur--Weyl multiplicities $5,3,1$ we find that
\[
\begin{aligned}
\HH_2^4
&\downarrow
5A_1
\;\oplus\;
3(A_1 \oplus B_1 \oplus B_2)
\;\oplus\;
(A_1 \oplus A_2)\\[3pt]
&\cong
9A_1 \;\oplus\; 1A_2 \;\oplus\; 3B_1 \;\oplus\; 3B_2.
\end{aligned}
\]
Hence \ce{SF4} exhibits four nuclear--spin species, with statistical weights in
the ratio
\[
A_1:A_2:B_1:B_2 \;=\; 9:1:3:3,
\]
which can be verified using standard methods~\cite{BunkerJensen}.

As in the previous examples, this analysis can be refined by total spin.  Using
\[
(4)\leftrightarrow I=2,\qquad
(3,1)\leftrightarrow I=1,\qquad
(2,2)\leftrightarrow I=0,
\]
we obtain the refined, spin--resolved contributions
\[
I=2:\ 5A_1,
\;
I=1:\ 3(A_1\oplus B_1\oplus B_2),
\;
I=0:\ (A_1\oplus A_2).
\]

This example illustrates that cyclic point--group symmetries arising from
molecular geometries may embed $C_m$ into $S_N$ either as a single $N$--cycle or
as a product of disjoint $m$--cycles.  In both cases, the major--index branching
rule provides a uniform, purely combinatorial method—based entirely on Young
diagrams and standard tableaux—for passing from the full permutation symmetry
$S_N$ to the physically realized cyclic subgroup and hence for computing
nuclear--spin statistical weights.

\begin{table*}[ht]
\centering
\caption{Nuclear--spin statistical weights for each molecule considered in this work.
For each example we list the identical nuclei, their spin $I$, the number $N$ of
identical nuclei, the physically realized permutation subgroup of $S_N$, the
molecular point group, and the resulting statistical weights.}
\begin{tabular}{cccccc}
\hline
Molecule & Nuclei & $I$ & Perm.\ subgroup & Point group & Statistical weights \\
\hline
\ce{XeOF4}   & F  & $\tfrac{1}{2}$ & $C_4 \subset S_4$ & $C_{4v}$ & $A_1:A_2:B_1:B_2:E = 6:0:3:1:3$ \\[4pt]
\ce{SF4}     & F  & $\tfrac{1}{2}$ & $C_2 \subset S_4$ & $C_{2v}$ & $A_1:A_2:B_1:B_2 = 9:1:3:3$ \\[4pt]
\ce{C6H6}    & H  & $\tfrac{1}{2}$ & $C_6 \subset S_6$ & $D_{6h}$ & $A_{1g}:A_{2g}:B_{1g}:B_{2g}:E_{1g}:E_{2g} = 13:1:7:3:9:11$ \\[4pt]
\ce{C6D6}    & D  & $1$            & $C_6 \subset S_6$ & $D_{6h}$ & $A_1:A_2:B_1:B_2:E_1:E_2 = 92:38:73:46:116:124$ \\[4pt]
\ce{C7H7^+}  & H  & $\tfrac{1}{2}$ & $C_7 \subset S_7$ & $D_{7h}$ & $A_1:A_2:E_1:E_2:E_3 = 18:2:18:18:18$ \\
\hline
\end{tabular}
\label{tab:summary}
\end{table*}

\section{Conclusion}

The Schur--Weyl perspective introduced into molecular physics by Schmiedt, Jensen,
and Schlemmer~\cite{SchmiedtJensenSchlemmer2016} revealed that nuclear--spin permutation
symmetry and collective nuclear--spin structure arise from a single
representation--theoretic decomposition of the $n$--spin Hilbert space.
In the present work we have continued this program by treating the experimentally
relevant situation in which the rovibrational geometry realizes only a subgroup of the
full permutation group, most commonly a cyclic subgroup generated by an $m$--fold
rotation, and by explicitly incorporating the corresponding dihedral point--group
symmetry where reflections are present. The results of these investigations are presented in Table~\ref{tab:summary}.

By combining the Schur--Weyl decomposition with the
Kraśkiewicz--Weyman--Adin--Roichman major--index branching rule for the restriction
$S_N \downarrow C_m$, we obtain a direct and fully combinatorial procedure for computing
nuclear--spin statistical weights.
Each standard Young tableau appearing in the Schur--Weyl decomposition carries a definite
$C_m$ species determined by its major index modulo $m$, and the subsequent action of
reflections assigns the corresponding dihedral symmetry.
The nuclear--spin symmetry content therefore follows directly from tableau data, without
the use of projection operators, cycle--index polynomials, or case--specific symmetry
constructions.

Applications to \ce{XeOF4}, \ce{SF4},  benzene, deuterated benzene and the tropylium cation show
that the method applies uniformly to realistic molecular geometries, including cases in
which the permutation symmetry is generated by a single cycle or by a product of disjoint
cycles.
In each case, the resulting spin--resolved nuclear--spin species and statistical weights
agree with established results while providing a clearer organizational structure and a
straightforward route to automation.
The method directly supplies the spin--weight factors required for rovibrational line
lists and intensity modeling.

From a spectroscopic perspective, these statistical weights provide the
symmetry-dependent factors attached to rovibrational manifolds and hence enter the
interpretation of relative intensities among symmetry-distinct rotational and
rovibrational transitions.  The examples considered here also show that the outcome is
controlled both by the permutation subgroup realized by the molecular geometry and by
the single-nucleus spin $I$: molecules with the same geometric symmetry but different
nuclear spin can exhibit genuinely different Young-diagram content and therefore
different statistical-weight patterns.  In this sense, the present framework connects
observable spectroscopic consequences directly to a uniform representation-theoretic
description of nuclear-spin symmetry.

Conceptually, the framework reinforces the unifying viewpoint of
Ref.~\cite{SchmiedtJensenSchlemmer2016}.  Young diagrams and tableaux encode both
permutation symmetry and collective nuclear spin within a single representation, and
restriction to the physically realized permutation subgroup selects the symmetry species
relevant for spectroscopy.  Because Schur--Weyl duality accommodates arbitrary
single--nucleus spin $I$, the present construction extends naturally to higher--spin
isotopologues and to systems with larger numbers of identical nuclei.  By continuing
the Schur--Weyl program in a form adapted to molecular point--group symmetries actually
realized in experiment, this work provides both conceptual clarity and practical
computational tools for the analysis of nuclear--spin symmetry in polyatomic molecules.

\section*{Acknowledgements}

This research was supported in part by the NSF Quantum Leap Challenge Institute grant OMA-2120757 and the MathQuantum RTG through the NSF RTG grant DMS-2231533. All figures are obtained from Wikimedia Commons and are used in accordance with their respective licenses.

\bibliographystyle{apsrev4-2}
\bibliography{biblio}

\begin{thebibliography}{10}%
\makeatletter
\providecommand \@ifxundefined [1]{%
 \@ifx{#1\undefined}
}%
\providecommand \@ifnum [1]{%
 \ifnum #1\expandafter \@firstoftwo
 \else \expandafter \@secondoftwo
 \fi
}%
\providecommand \@ifx [1]{%
 \ifx #1\expandafter \@firstoftwo
 \else \expandafter \@secondoftwo
 \fi
}%
\providecommand \natexlab [1]{#1}%
\providecommand \enquote  [1]{``#1''}%
\providecommand \bibnamefont  [1]{#1}%
\providecommand \bibfnamefont [1]{#1}%
\providecommand \citenamefont [1]{#1}%
\providecommand \href@noop [0]{\@secondoftwo}%
\providecommand \href [0]{\begingroup \@sanitize@url \@href}%
\providecommand \@href[1]{\@@startlink{#1}\@@href}%
\providecommand \@@href[1]{\endgroup#1\@@endlink}%
\providecommand \@sanitize@url [0]{\catcode `\\12\catcode `\$12\catcode
  `\&12\catcode `\#12\catcode `\^12\catcode `\_12\catcode `\%12\relax}%
\providecommand \@@startlink[1]{}%
\providecommand \@@endlink[0]{}%
\providecommand \url  [0]{\begingroup\@sanitize@url \@url }%
\providecommand \@url [1]{\endgroup\@href {#1}{\urlprefix }}%
\providecommand \urlprefix  [0]{URL }%
\providecommand \Eprint [0]{\href }%
\providecommand \doibase [0]{https://doi.org/}%
\providecommand \selectlanguage [0]{\@gobble}%
\providecommand \bibinfo  [0]{\@secondoftwo}%
\providecommand \bibfield  [0]{\@secondoftwo}%
\providecommand \translation [1]{[#1]}%
\providecommand \BibitemOpen [0]{}%
\providecommand \bibitemStop [0]{}%
\providecommand \bibitemNoStop [0]{.\EOS\space}%
\providecommand \EOS [0]{\spacefactor3000\relax}%
\providecommand \BibitemShut  [1]{\csname bibitem#1\endcsname}%
\let\auto@bib@innerbib\@empty
\bibitem [{\citenamefont {Schmiedt}\ \emph {et~al.}(2016)\citenamefont
  {Schmiedt}, \citenamefont {Jensen},\ and\ \citenamefont
  {Schlemmer}}]{SchmiedtJensenSchlemmer2016}%
  \BibitemOpen
  \bibfield  {author} {\bibinfo {author} {\bibfnamefont {H.}~\bibnamefont
  {Schmiedt}}, \bibinfo {author} {\bibfnamefont {P.}~\bibnamefont {Jensen}},\
  and\ \bibinfo {author} {\bibfnamefont {S.}~\bibnamefont {Schlemmer}},\ }\href
  {https://doi.org/10.1063/1.4960956} {\bibfield  {journal} {\bibinfo
  {journal} {The Journal of Chemical Physics}\ }\textbf {\bibinfo {volume}
  {145}},\ \bibinfo {pages} {074301} (\bibinfo {year} {2016})}\BibitemShut
  {NoStop}%
\bibitem [{\citenamefont {Fulton}\ and\ \citenamefont
  {Harris}(1999)}]{fultonharris}%
  \BibitemOpen
  \bibfield  {author} {\bibinfo {author} {\bibfnamefont {W.}~\bibnamefont
  {Fulton}}\ and\ \bibinfo {author} {\bibfnamefont {J.}~\bibnamefont
  {Harris}},\ }\href@noop {} {\emph {\bibinfo {title} {Representation
  theory}}},\ \bibinfo {edition} {1st}\ ed.,\ Graduate texts in mathematics\
  (\bibinfo  {publisher} {Springer},\ \bibinfo {address} {New York, NY},\
  \bibinfo {year} {1999})\BibitemShut {NoStop}%
\bibitem [{\citenamefont {Longuet-Higgins}(1963)}]{LonguetHiggins1963NonRigid}%
  \BibitemOpen
  \bibfield  {author} {\bibinfo {author} {\bibfnamefont {H.~C.}\ \bibnamefont
  {Longuet-Higgins}},\ }\href {https://doi.org/10.1080/00268976300100501}
  {\bibfield  {journal} {\bibinfo  {journal} {Molecular Physics}\ }\textbf
  {\bibinfo {volume} {6}},\ \bibinfo {pages} {445} (\bibinfo {year}
  {1963})}\BibitemShut {NoStop}%
\bibitem [{\citenamefont {Quack}(1977)}]{Quack1977SymmetrySelection}%
  \BibitemOpen
  \bibfield  {author} {\bibinfo {author} {\bibfnamefont {M.}~\bibnamefont
  {Quack}},\ }\href {https://doi.org/10.1080/00268977700101861} {\bibfield
  {journal} {\bibinfo  {journal} {Molecular Physics}\ }\textbf {\bibinfo
  {volume} {34}},\ \bibinfo {pages} {477} (\bibinfo {year} {1977})}\BibitemShut
  {NoStop}%
\bibitem [{\citenamefont {Park}\ and\ \citenamefont
  {Light}(2007)}]{ParkLight2007SpinSelection}%
  \BibitemOpen
  \bibfield  {author} {\bibinfo {author} {\bibfnamefont {K.}~\bibnamefont
  {Park}}\ and\ \bibinfo {author} {\bibfnamefont {J.~C.}\ \bibnamefont
  {Light}},\ }\href {https://doi.org/10.1063/1.2805394} {\bibfield  {journal}
  {\bibinfo  {journal} {The Journal of Chemical Physics}\ }\textbf {\bibinfo
  {volume} {127}},\ \bibinfo {pages} {224101} (\bibinfo {year}
  {2007})}\BibitemShut {NoStop}%
\bibitem [{\citenamefont {Oka}(2004)}]{Oka2004NuclearSpinSelection}%
  \BibitemOpen
  \bibfield  {author} {\bibinfo {author} {\bibfnamefont {T.}~\bibnamefont
  {Oka}},\ }\href {https://doi.org/10.1016/j.jms.2004.08.015} {\bibfield
  {journal} {\bibinfo  {journal} {Journal of Molecular Spectroscopy}\ }\textbf
  {\bibinfo {volume} {228}},\ \bibinfo {pages} {635} (\bibinfo {year}
  {2004})}\BibitemShut {NoStop}%
\bibitem [{\citenamefont {Bengs}(2020)}]{Bengs2020}%
  \BibitemOpen
  \bibfield  {author} {\bibinfo {author} {\bibfnamefont {C.}~\bibnamefont
  {Bengs}},\ }\href {https://doi.org/10.1063/1.5140186} {\bibfield  {journal}
  {\bibinfo  {journal} {The Journal of Chemical Physics}\ }\textbf {\bibinfo
  {volume} {152}},\ \bibinfo {pages} {054106} (\bibinfo {year}
  {2020})}\BibitemShut {NoStop}%
\bibitem [{\citenamefont {Kr\'askiewicz}\ and\ \citenamefont
  {Weyman}(2001)}]{KraskiewiczWeyman1989}%
  \BibitemOpen
  \bibfield  {author} {\bibinfo {author} {\bibfnamefont {W.}~\bibnamefont
  {Kr\'askiewicz}}\ and\ \bibinfo {author} {\bibfnamefont {J.}~\bibnamefont
  {Weyman}},\ }\href@noop {} {\bibfield  {journal} {\bibinfo  {journal}
  {Bayreuth. Math. Schr.}\ ,\ \bibinfo {pages} {265}} (\bibinfo {year}
  {2001})},\ \bibinfo {note} {originally circulated 1989}\BibitemShut {NoStop}%
\bibitem [{\citenamefont {Adin}\ and\ \citenamefont
  {Roichman}(2001)}]{AdinRoichman2001}%
  \BibitemOpen
  \bibfield  {author} {\bibinfo {author} {\bibfnamefont {R.~M.}\ \bibnamefont
  {Adin}}\ and\ \bibinfo {author} {\bibfnamefont {Y.}~\bibnamefont
  {Roichman}},\ }\href {https://doi.org/10.1006/eujc.2000.0469} {\bibfield
  {journal} {\bibinfo  {journal} {European Journal of Combinatorics}\ }\textbf
  {\bibinfo {volume} {22}},\ \bibinfo {pages} {431} (\bibinfo {year}
  {2001})}\BibitemShut {NoStop}%
\bibitem [{\citenamefont {Bunker}\ and\ \citenamefont
  {Jensen}(2006)}]{BunkerJensen}%
  \BibitemOpen
  \bibfield  {author} {\bibinfo {author} {\bibfnamefont {P.~R.}\ \bibnamefont
  {Bunker}}\ and\ \bibinfo {author} {\bibfnamefont {P.}~\bibnamefont
  {Jensen}},\ }\href@noop {} {\emph {\bibinfo {title} {Fundamentals of
  Molecular Symmetry}}},\ \bibinfo {edition} {2nd}\ ed.\ (\bibinfo  {publisher}
  {CRC Press},\ \bibinfo {year} {2006})\BibitemShut {NoStop}%
\end{thebibliography}%

\end{document}